\documentclass[preprintnumbers,secnumarabic]{revtex4}
\pdfoutput=1
\usepackage{graphicx}
\usepackage{amsmath,amssymb,mathrsfs}
\usepackage{fancyhdr}
\usepackage{array,bbm} 

\newcommand*{\trans}{\mathrm{T}}                     
\newcommand*{\unitmatrix}{\mathbbm{1}}
\newcommand*{\tvec}[1]{\ensuremath{\boldsymbol{\mathrm{#1}}}}           

\makeatletter       

\renewcommand{\p@subsection}{}

\makeatother

\DeclareMathOperator{\diag}{diag}

\begin{document}

\preprint{HD-THEP-10-16}

\title{On the phenomenology of a two-Higgs-doublet model with
maximal CP symmetry\\ at the LHC -- synopsis and addendum}

\author{M. Maniatis}
    \email[E-mail: ]{M.Maniatis@thphys.uni-heidelberg.de}
    \affiliation{
Institut f\"ur Theoretische Physik, 
Universit\"at~Heidelberg,
Philosophenweg~16, 
69120~Heidelberg, 
Germany
}
\author{A.~von Manteuffel}
    \email[E-mail: ]{Manteuffel@physik.uzh.ch}
    \affiliation{
    Institut~f\"ur~Theoretische~Physik, 
Universit\"at Z\"urich, 
Winterthurerstr.~190, 
8057~Z\"urich,
Switzerland
}
\author{O. Nachtmann}
    \email[E-mail: ]{O.Nachtmann@thphys.uni-heidelberg.de}
    \affiliation{
Institut f\"ur Theoretische Physik, 
Universit\"at~Heidelberg,
Philosophenweg~16, 69120~Heidelberg, 
Germany
}

\begin{abstract}
Predictions for LHC physics are given for a two-Higgs-doublet model
having four generalized CP symmetries. 
In this
{\em maximally-CP-symmetric model}
(MCPM) the first fermion family is, at tree level, uncoupled to the Higgs fields and
thus massless. The second and third fermion families have a very symmetric coupling
to the Higgs fields. But through the electroweak symmetry breaking a large mass
hierarchy is generated between these fermion families. Thus, the fermion mass spectrum
of the model presents a rough approximation to what is observed in Nature.
In the MCPM the couplings of the Higgs bosons to the fermions
are completely fixed. This allows us to present clear predictions
for the production at the LHC and for the decays of the physical Higgs bosons.
As salient feature we find rather large cross sections for Higgs-boson
production via Drell--Yan type processes.
In this paper we present a short outline of the model
and extend a former study by the
predictions at LHC for a center-of-mass energy of 7~TeV. 
\end{abstract}

\maketitle

\section{introduction}

Extending the Standard Model~(SM) Higgs sector to two Higgs doublets,
\begin{equation}
\varphi_1=
\begin{pmatrix} \varphi^+_1\\ \varphi^0_1 \end{pmatrix},
\qquad
\varphi_2=
\begin{pmatrix} \varphi^+_2\\ \varphi^0_2 \end{pmatrix} \,,
\end{equation}
gives the two-Higgs-doublet model~(THDM).
There, the potential may contain many
more terms than in the SM; see e.g.~\cite{Gunion:1989we,Gunion:1992hs}.
The most general THDM Higgs potential 
can be written~\cite{Haber:1993an}
\begin{multline}
\label{V_fields}
V =
m_{11}^2 (\varphi_1^\dagger \varphi_1) +
m_{22}^2 (\varphi_2^\dagger \varphi_2) -
m_{12}^2 (\varphi_1^\dagger \varphi_2) -
(m_{12}^2)^* (\varphi_2^\dagger \varphi_1)
+\frac{1}{2} \lambda_1 (\varphi_1^\dagger \varphi_1)^2
+ \frac{1}{2} \lambda_2 (\varphi_2^\dagger \varphi_2)^2
+ \lambda_3 (\varphi_1^\dagger \varphi_1)(\varphi_2^\dagger \varphi_2)
\\
+ \lambda_4 (\varphi_1^\dagger \varphi_2)(\varphi_2^\dagger \varphi_1)
+ \frac{1}{2} [\lambda_5 (\varphi_1^\dagger \varphi_2)^2 + \lambda_5^*
(\varphi_2^\dagger \varphi_1)^2]
+ [\lambda_6 (\varphi_1^\dagger \varphi_2) + \lambda_6^*
(\varphi_2^\dagger \varphi_1)] (\varphi_1^\dagger \varphi_1) + [\lambda_7 (\varphi_1^\dagger
\varphi_2) + \lambda_7^* (\varphi_2^\dagger \varphi_1)] (\varphi_2^\dagger \varphi_2)\;,
\end{multline}
with $m_{11}^2$, $m_{22}^2$, $\lambda_{1,2,3,4}$ real and
$m_{12}^2$, $\lambda_{5,6,7}$ complex.
Many properties of THDMs turn out 
to have a simple geometric meaning if we introduce
gauge invariant
bilinears~\cite{Nagel:2004sw,Maniatis:2006fs},
\begin{equation}
K_0 = \varphi_1^{\dagger} \varphi_1 + \varphi_2^{\dagger} \varphi_2, \quad
\tvec{K} =
\begin{pmatrix}
K_1 \\ K_2 \\ K_3
\end{pmatrix}
=
\begin{pmatrix}
\varphi_1^\dagger \varphi_2 + \varphi_2^\dagger \varphi_1\\
i \varphi_2^\dagger \varphi_1 - i \varphi_1^\dagger \varphi_2\\
\varphi_1^{\dagger} \varphi_1 - \varphi_2^{\dagger} \varphi_2
\end{pmatrix}.
\end{equation}
In terms of these bilinears $K_0$, $\tvec{K}$, the 
Higgs potential~\eqref{V_fields} reads
\begin{equation}
V = \xi_0 K_0 + \tvec{\xi}^\trans \tvec{K}
  + \eta_{00} K_0^2
  + 2 K_0\tvec{\eta}^\trans \tvec{K}
  + \tvec{K}^\trans E \tvec{K}
\end{equation}
with parameters~$\xi_0$, $\eta_{00}$,
3-component vectors $\tvec{\xi}$, $\tvec{\eta}$
and a $3 \times 3$ matrix $E=E^\trans$, all real.

The standard CP transformation of the Higgs-doublet fields is defined by
\begin{equation}
\label{eq-simCP}
\varphi_i(x) \rightarrow \phantom{-}\varphi_i^*(x')\,, \qquad i=1,2\,\,
\qquad x'=(x^0, -\tvec{x}).
\end{equation}
In terms of the bilinears, this standard CP~transformation 
is~\cite{Nishi:2006tg,Maniatis:2007vn}
\begin{equation}
\label{eq-simCPK}
K_0(x) \rightarrow K_0(x')\;,\quad
\tvec{K}(x) \rightarrow \bar{R}_2 \tvec{K}(x')
\end{equation}
where $\bar{R}_2=\diag(1,-1,1)$, corresponding in $K$~space
to a reflection 
on the 1--3 plane.
Generalised CP transformations~(GCPs) are \mbox{defined} by
\cite{Lee:1966ik,Ecker:1981wv,Ecker:1983hz}
\begin{equation}
\varphi_i(x) \rightarrow U_{ij} \; \varphi_j^*(x'),
\quad i,j=1,2\,,
\end{equation}
with $U$ an arbitrary unitary $2 \times 2$ matrix.
In terms of the bilinears this reads~\cite{Maniatis:2007vn}
\begin{equation}
K_0(x)  \rightarrow K_0(x'),\quad
\tvec{K}(x) \rightarrow \bar{R}\; \tvec{K}(x')
\end{equation}
with an improper rotation matrix~$\bar{R}$.

Requiring $\bar{R}^2=\unitmatrix_3$ leads to two types
of GCPs. In $K$ space:
\begin{alignat*}{2}
&(i)\phantom{i} \quad \bar{R}=-\unitmatrix_3, \quad &&\text{point reflection,}\\
&(ii) \quad \bar{R}= R^\trans \; \bar{R}_2\; R,\quad &&
\text{reflection on a plane}~(R \in SO(3) ).
\end{alignat*}

While the CP transformations of type $(ii)$ are 
equivalent to the standard CP transformation~\eqref{eq-simCP}, 
respectively~\eqref{eq-simCPK}, the point reflection transformation
of type $(i)$ is quite different and turns out
to  have very interesting properties.
Motivated by this geometric picture of
generalised CP transformations,
the most general THDM invariant under the point reflection~($i$) has
been studied in~\cite{Maniatis:2007de,Maniatis:2009vp,Maniatis:2009by}.
The corresponding potential has to obey the conditions
$\tvec{\xi}=\tvec{\eta}=0$,
\begin{equation}
V_{\text{MCPM}} =
  \xi_0\, K_0 
  + \eta_{00}\, K_0^2
  + \tvec{K}^\trans\, E\, \tvec{K}\,.
\end{equation}
This model is, besides the point reflection symmetry of type $(i)$, 
invariant under three GCPs
of type $(ii)$.
We call this model therefore maximally CP symmetric model, MCPM.
Requiring also maximally CP symmetric 
Yukawa couplings
we find that at least two fermion families are 
necessary in order to have non-vanishing 
fermion masses. That is, we find a reason for
family replication in the MCPM.
Furthermore, requiring absence of large flavor changing
neutral currents it was shown that
the Yukawa couplings are completely fixed.
For instance for the lepton sector we get the Yukawa couplings
\begin{equation}
\mathscr{L}_{\mathrm{Yuk}} =
 - 
\sqrt{2}\frac{m_\tau}{v}
\bigg\{
   \bar{\tau}_{R}\,\varphi_1^\dagger
   \begin{pmatrix} \nu_{\tau} \\ \tau \end{pmatrix}_{\!\!\! L}
  -
   {
   \bar{\mu}_{R}\,\varphi_2^\dagger
   \begin{pmatrix} \nu_{\mu} \\ \mu \end{pmatrix}_{\!\!\! L}
   }             \bigg\}
        + h.c.
\end{equation}
In the unitary gauge electroweak symmetry breaking gives
\begin{equation}
\varphi_1 =
\frac{1}{\sqrt{2}}
\begin{pmatrix}
0 \\ v_0 + \rho'
\end{pmatrix}\;,
\quad
\varphi_2 =
\begin{pmatrix}
H^+ \\
\frac{1}{\sqrt{2}} ( h' + i h'' )
\end{pmatrix}
\end{equation}
with the standard vacuum-expectation value \mbox{$v_0 \approx 246$~GeV}.
The physical Higgs-boson fields are
$\rho'$, $h'$, $h''$, and $H^\pm$.

Let us briefly summarize the essential
properties of the MCPM:
\begin{itemize}
\item There are 5 physical Higgs particles, two CP even ones $\rho'$, $h'$, 
one CP odd one $h''$, and a charged Higgs-boson pair~$H^\pm$.
\item The $\rho'$ boson couples exclusively to the third $(\tau, t, b)$~family,
$\rho'$ behaves like the SM Higgs boson.
\item The Higgs bosons $h'$, $h''$, $H^\pm$ couple exclusively to
the second $(\mu, c, s)$~family with
strengths proportional to the masses of the third generation fermions.
\item The first $(e, u, d)$~family is uncoupled to the Higgs bosons.
\end{itemize}
For further details
we refer to~\cite{Maniatis:2007de}.


\section{Predictions for hadron colliders}

\begin{center}
\begin{figure}[t]
\begin{tabular}{m{0.4\linewidth}m{0.4\linewidth}}
\centerline{\includegraphics[bb=282 170 760 417,width=0.3\textwidth,clip]{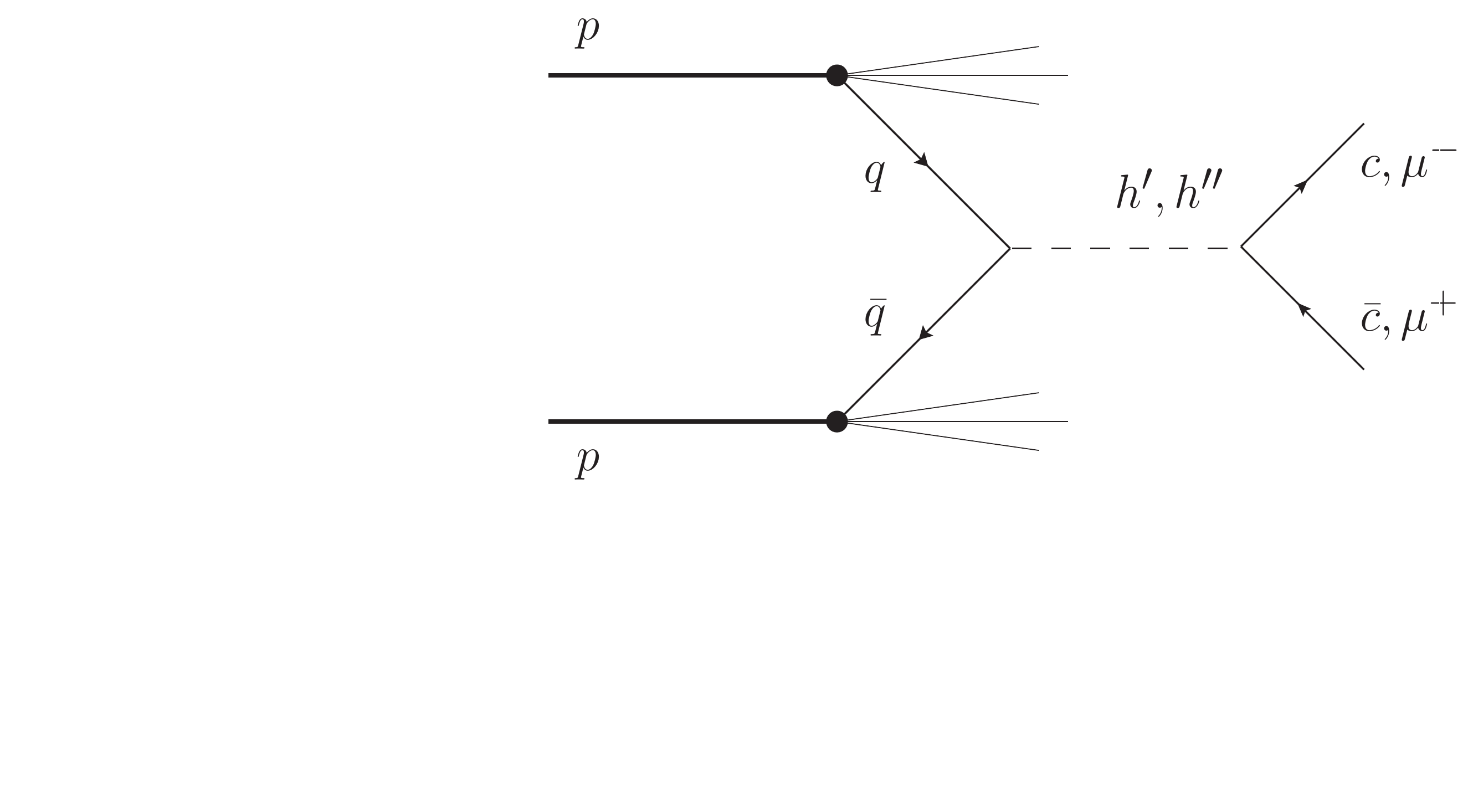}} &
\centerline{\includegraphics[bb=282 170 760 417,width=0.3\textwidth,clip]{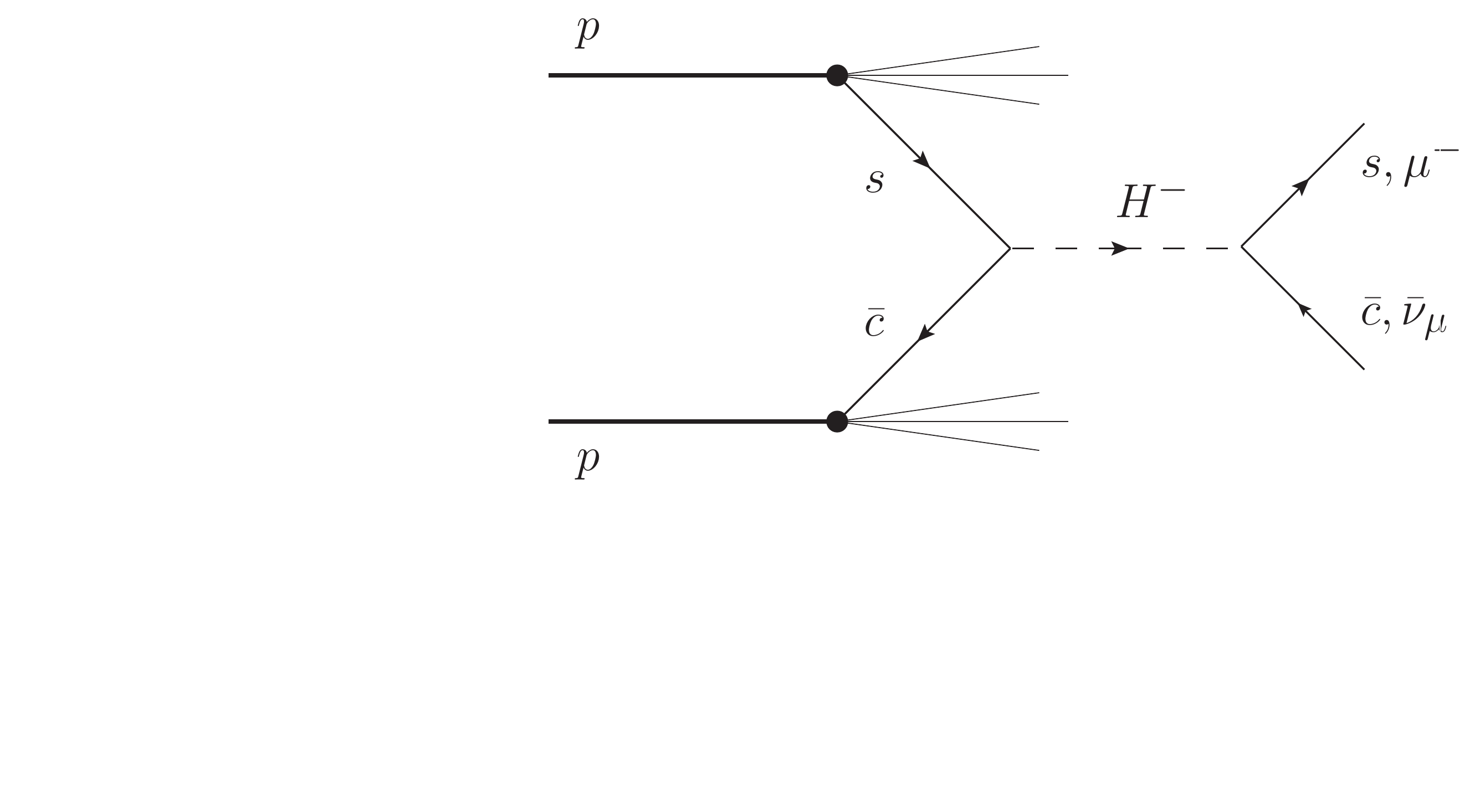}}
\end{tabular}
\caption{ \label{fig-diag}
Feynman diagrams for the Drell--Yan type 
Higgs-boson production and decay reactions
which are enhanced in the MCPM ($q=c,s$).}
\end{figure}
\end{center}

\begin{figure}[t]
\includegraphics[width=0.45\textwidth,clip]{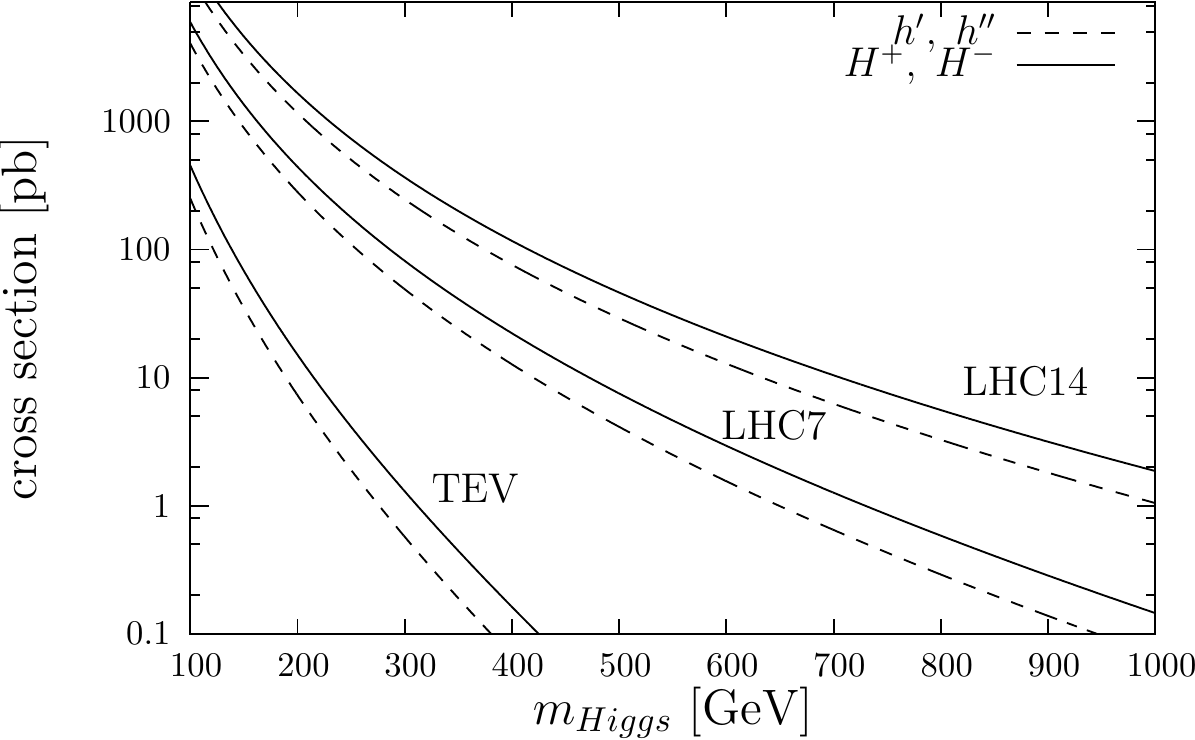} \qquad
\includegraphics[width=0.45\textwidth,clip]{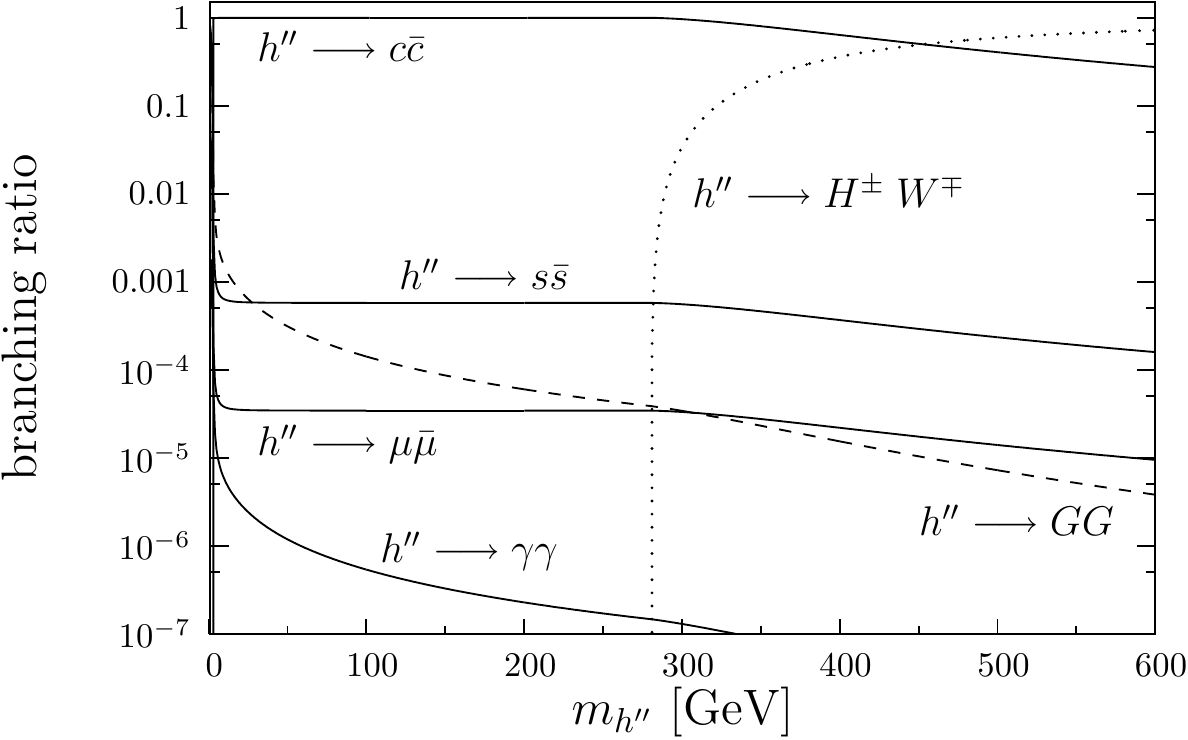}
\caption{\label{figs} left: total cross section of Drell--Yan
type Higgs boson production at TEVATRON
and LHC.
right: branching ratios of the CP odd $h''$ Higgs boson,
where a mass of $m_{H^\pm}=200$~GeV is assumed.}
\end{figure}

Since the Yukawa couplings of the
$h'$, $h''$, $H^\pm$ Higgs bosons to
the second fermion family are proportional
to the third-fermion-family masses we have large 
cross sections for Drell--Yan type Higgs-boson production.
For the same reason we have large decay rates of
these Higgs bosons to the second generation fermions.
In figure~\ref{fig-diag} we show the diagrams
for these production and decay reactions in $pp$~collisions.
In~\cite{Maniatis:2009vp}
the cross sections were computed for Drell-Yan Higgs-boson 
production at the TEVATRON and the LHC for center-of-mass
energies of 1.96~TeV and 14~TeV, respectively.
In~\cite{Maniatis:2009by} radiative effects were considered.
Here we add the cross sections for a center-of-mass energy
of~7~TeV at LHC, which is currently available.
The corresponding total cross sections for the Drell--Yan 
production of the 
$h'$, $h''$, $H^\pm$ bosons
are shown in figure~\ref{figs}.
In this figure we also recall the
branching ratios of the $h''$ boson decays.
As an example consider
Higgs-boson masses $h'$, $h''$, $H^\pm$ of 200~GeV
where we get very large total production cross sections,
around 850~pb, for LHC7.
These Higgs bosons
decay mainly into light $c$ and $s$ quarks.
However, tagging of $c$ and $s$-quarks
in the detectors is at least challenging.
Channels involving muons should be more easily
accessible experimentally.
With the branching ratio of $3 \times 10^{-5}$
into $\mu$-pairs, we predict about 25 $\mu$ events from 
a 200~GeV $h'$~($h''$)
at LHC7~for~1~fb$^{-1}$ integrated luminosity. For further details of
the calculations we refer to~\cite{Maniatis:2007de,Maniatis:2009vp,Maniatis:2009by}.


\end{document}